# Are Attributions of Consciousness to AI Chatbots Epistemically Innocent?

Uwe Peters
Utrecht University

## Abstract

Artificial intelligence (AI) chatbots (e.g., ChatGPT) can communicate in strikingly humanlike ways. This has prompted many chatbot users to attribute psychological properties, including consciousness, to these systems. However, there is little scientific evidence that current AI chatbots are conscious. How, then, should we understand people's consciousness attributions to chatbots? Are they merely metaphorical claims, or do they express genuine beliefs? If these attributions lack evidential support, are users epistemically blameworthy for making them, or might they be epistemically innocent, yielding significant benefits otherwise unattainable? This paper offers a conceptual analysis of consciousness attributions to AI chatbots and develops a multidimensional taxonomy of the attitudes they may express, ranging from non-doxastic stances (e.g., pretence) to different forms of belief, including delusions. This taxonomy helps avoid conflations by showing that linguistically identical attributions can reflect importantly different attitudes and degrees of epistemic commitment to the proposition that chatbots are conscious. The taxonomy also provides a framework for empirical studies to operationalize and measure different forms of epistemic commitment to AI consciousness. Using this taxonomy, I argue that although some consciousness attributions to chatbots are epistemically benign, and even some irrational ones may be epistemically innocent, many others render the attributor epistemically blameworthy.

Keywords: AI chatbots; consciousness attribution; attitudes; epistemic benefits

## 1. Introduction

ChatGPT and other AI chatbots based on large language models (LLMs) can produce compellingly humanlike outputs, leading many chatbot users to ascribe psychological properties to these systems (Reinicke et al., 2025). In fact, recent surveys found that the majority of participants claimed that ChatGPT was conscious, i.e., had subjective experience (Colombatto & Fleming, 2024; Kang et al., 2025). Given the increasingly anthropomorphic design and sophistication of AI chatbots (including social AI companions, e.g., Replika), people's consciousness attributions to these systems are likely to increase soon (Shevlin, 2024).

However, most experts in AI and consciousness science hold that there is little or no evidence that any current AI chatbot is conscious (Suleyman, 2025; Dreksler et al., 2025; Seth, 2026),[1] with some contributors writing that the "general consensus is that LLMs are not conscious" (Prettyman, 2024, p. 1). Claims attributing consciousness to current AI chatbots therefore appear to go beyond the available evidence and expert consensus (McClelland, 2025).

Given this, how are we to understand these consciousness attributions? For instance, previous philosophical contributions have noted that they might be merely figurative statements (Coghlan, 2024), be literal ("unironic") expressions of a genuine belief, or be indicative of a variety of other attitudes situated between make-believe and full belief (Shevlin, 2024).

[1] https://mustafa-suleyman.ai/seemingly-conscious-ai-is-coming

Determining what mental states these attributions express matters because if they indicate genuine beliefs, not just verbal endorsements, this may reveal overestimations of chatbot capacities, raising risks of people overtrusting these systems. However, the potential variety of the mental states expressed by consciousness attributions to AI chatbots hasn't been systematically investigated yet.

Moreover, it remains unexplored whether some or all of these attributions are epistemically rational or irrational, even delusional. For instance, after building ELIZA, which is thought to be the first chatbot, and witnessing that users readily ascribed understanding to it, Joseph Weizenbaum (1976) viewed this as "powerful delusional thinking" (p. 7). Relatedly, in the news and on social media, certain kinds of consciousness attributions to AI chatbots have recently been called instances of "AI psychosis," a phenomenon in which chatbots amplify, validate, or co-produce convictions that something imaginary has become real (Morrin et al., 2025). It is currently unclear whether, which, or when consciousness attributions to AI chatbots are epistemically rational or delusional.

Similarly, it remains largely unexplored how we should respond to people producing such attributions. Since experts widely agree that current AI chatbots aren't conscious (Dreksler et al., 2025; Seth, 2026) or that the evidence remains inconclusive (McClelland, 2025), are people epistemically blameworthy when they nonetheless attribute consciousness to such systems? Should we try to correct them for their attributions?

To make progress on these questions, my goal here is twofold. First, I will analyze consciousness attributions to AI chatbots and taxonomize the underlying states. The taxonomy is intended to clarify the conceptual space of consciousness attributions to chatbots so as to reduce the risk of conflating different states that may manifest in linguistically identical attributions and provide a framework for future empirical studies to operationalize and test for different forms of epistemic commitment to chatbot consciousness.

Second, putting this taxonomy to work and drawing on research on "epistemic innocence" of imperfect cognitions (Bortolotti, 2020), I argue that some consciousness attributions to AI chatbots are epistemically benign for the attributor, as they don't involve evidence-recalcitrant states, and even some epistemically irrational ones are epistemically innocent in that they can have significant epistemic benefits unattainable otherwise. However, many others render the attributor epistemically blameworthy.

Much of the following discussion can be applied to anthropomorphic attributions to AI more generally. But the focus here will remain on consciousness attributions because they can be especially consequential. The reason is that even though consciousness is distinct from sentience, i.e., positively or negatively valanced experience (pleasure, pain, etc.),[2] the two are often conflated (Birch, 2024). This matters ethically, as sentience is commonly thought to be sufficient for moral status (for discussion, see Shepherd, 2024). If people are convinced that AI chatbots are conscious and equate this with them having valenced experience, they may hold that these systems also deserve protection from harm, which could have profound legal and policy implications (Caviola et al., 2025), potentially fuelling calls for "AI welfare" (Goldstein & Kirk-Giannini, 2026) (for scepticism about AI welfare, see Dorsch et al., 2025).

[2] One may subjectively experience, say, a white wall without the experience being either positively or negatively valenced.

## 2. Consciousness attributions to AI chatbots: A taxonomy

A first key distinction when assessing attributions of psychological features to beings or things in general is that between a nonverbal or verbal *expression* (e.g., a sentence) and a *thought*. This is because by 'attribution' one may mean, for instance, a sentence (e.g., "ChatGPT is conscious"), a thought, or both. While recent studies that explore consciousness attributions to AI chatbots take user self-reports to represent what respondents genuinely think or believe (e.g., Colombatto & Fleming, 2024; Dreksler et al., 2025), one might say "X is Y" without thinking or believing that X is Y. For clarity, I will here use 'attribution' to refer only to the nonverbal or verbal expression that may suggest the attributor thinks that X is Y.

A related distinction is between *private* and *public* consciousness attributions. A chatbot user may attribute consciousness to a chatbot only in inner speech, a personal journal, or private interactions with ChatGPT (Replika, etc.). Alternatively, the user may express it publicly (e.g., on Reddit), or both. Moreover, both public and private attributions may be either *direct* – explicitly ascribing consciousness (e.g., stating "ChatGPT is conscious") – or *indirect*, where consciousness is implied through the attribution of psychological predicates that commonly presuppose consciousness (e.g., "it cares") (Peter et al., 2025).

Finally, across contexts, consciousness attributions to chatbots may express one and the same or many different *attitudes* (Allport, 1935; Fodor, 1978), i.e., mental states relating an agent to the proposition that a chatbot is conscious and grounding the agent's attribution.[3] The question is what these attitudes are.

### 2.1 Methodological approach

Since consciousness attributions to AI chatbots are a relatively new phenomenon, there is little prior research to draw on when attempting to classify the underlying attitudes (Coghlan, 2024; Shevlin, 2024). Attempts to systematize this domain may thus risk appearing *ad hoc*.

However, more general questions of whether people genuinely believe what they report in surveys, what other attitudes than beliefs may underlie assertions seemingly indicating beliefs, and how to categorize evidentially unwarranted or even bizarre doxastic states (conspiracy theories, delusions, etc.) have been examined in political psychology (e.g., see the notion of "expressive responding," Schaffner & Luks, 2018), philosophy of mind (e.g., the notion of "in-between" belief; Schwitzgebel, 2001), epistemology (e.g., the notion of "acceptance" versus belief; Cohen, 1992; Hannon & De Ridder, 2021), and philosophy of psychiatry (e.g., Bortolotti, 2020). Drawing on this existing literature, we may distinguish at least three here relevant dimensions along which attitudes toward a proposition *p* can be characterized:

(a) *Commitment to the use of p*. This may be *practical commitment*, marking the extent to which one is disposed to select and express *p* in communication (e.g., within or across contexts) regardless of whether one treats *p* as true in one's own thinking (Schaffner & Luks, 2018), or *epistemic commitment*, marking the extent to which one treats *p* as true in inference, judgment, and decision-making. Epistemic commitment has two aspects, namely (i) whether *p* is taken up in one's inferential economy at all, and, if so, (ii) how stably it is maintained in the face of counterevidence.

[3] Some consciousness attributions to chatbots (e.g., in surveys) may not express any specific attitude but merely reflect random error, inattention, misunderstanding, or carelessness in the surveys (Ward & Meade, 2023). I shall set these cases aside here.

(b) *Doxastic status*. This relates to epistemic commitment and concerns whether the attitude involved is belief-like (doxastic) or not (e.g., accepting, supposing, regretting, etc. that *p*) (Cohen, 1992; Schwitzgebel, 2001).

(c) *Pathological status*. This concerns whether the attitude involved conforms to or deviates from standards of rationality and psychological functioning and may involve, for instance, distinguishing between delusional and non-delusional beliefs (e.g., Bortolotti, 2020).

I shall use these dimensions to analyze the attitudes that may underlie consciousness attributions to chatbots, thus extending existing concepts to this domain. This has the benefit that, while some of the categories introduced below are novel, none of them is posited *ad hoc* but constructed by adapting distinctions drawn elsewhere in the philosophical literature, while also integrating adjacent work from social psychology.

Against this background, at least the following ten attitudes may underlie consciousness attributions to chatbots:

(1) strategic pretence
(2) conformist affirmation
(3) allegiance affirmation
(4) imaginative affirmation
(5) acceptance
(6) in-between belief
(7) unbiased belief
(8) biased belief
(9) non-bizarre delusion
(10) bizarre delusion

I will explain (1)–(10) shortly. But to clarify, the list's lower to greater numbering isn't a one-dimensional scale but is meant to broadly track a multidimensional progression from (a) minimal to strong forms of practical and epistemic commitment to AI consciousness, (b) from non-doxastic to doxastic states, and (c) from non-pathological to pathological attitudes. These dimensions don't always vary independently or move in perfect lockstep, so the ordering should be read as approximate, not strictly linear.

Relatedly, the resulting taxonomy is hierarchical, not flat. Some categories mark different attitude-types (e.g., strategic pretence, imaginative affirmation, acceptance, belief). Others introduce refinements within a single type, especially belief. Thus, the later categories don't simply add more items to the list but increasingly differentiate doxastic states by their degree of inferential integration (e.g., from in-between to fully integrated belief), and resistance to counterevidence (low to high), culminating in pathological fixation (bizarre delusions).

Finally, these attitudes may be *explicit* (consciously accessible and measurable by self-report) or *implicit* (largely unconscious and testable through indirect measures (e.g., Implicit Association tasks (IAT), Greenwald et al., 2022)). The two can dissociate (Ajzen & Dasgupta, 2015): A user may sincerely explicitly deny that AI chatbots are conscious while nevertheless exhibiting nonverbal behavior (e.g., feeling guilt about 'hurting' a chatbot after insulting it, or reacting faster in IATs linking chatbots with consciousness) suggesting an opposing implicit attitude. Since empirical studies of AI consciousness attribution have relied exclusively on self-report, however, they have so far investigated only explicit attitudes (Colombatto & Fleming, 2024, p. 4). Accordingly, the taxonomy developed here focuses primarily on explicit attitudes.

### 2.2 From strategic pretence to delusions about conscious chatbots

This section elaborates on the ten attitudes just mentioned, beginning with cases in which attributors only practically commit themselves to the proposition that some AI chatbot is conscious without treating this proposition as true in their cognition.

(1) *Strategic pretence*

Some consciousness attributions to AI chatbots may involve no *epistemic* commitment to the corresponding proposition while still involving at least a weak practical commitment to select and communicate this proposition for instrumental reasons, for instance, to benefit from others coming to believe it. Consider cases in which AI companies that develop and commercialize chatbots describe their systems using psychological terms (e.g., "understands," "feels") to imply impressive model capabilities. For instance, the owners of Replika, a companion chatbot, market their chatbot as "emotionally intelligent," an "empathetic friend" (McStay, 2023) that "cares."[4] Since caring and empathy are widely understood as being able to "feel what someone else is feeling", positive or negative (Heyes, 2018, p. 499), such claims indirectly attribute sentience to the model. Yet, on a subpage, the company owner(s) also notes that "Replika is not a sentient being".[5]

This combination of suggestive indirect consciousness attribution and denial is characteristic of a stance I call *strategic pretence* – the attributor commits to choosing and (indirectly or directly) expressing the proposition that some AI system is conscious to attract attention and users, without treating it as true in cognition and while being ready to swiftly retract the claim. While the last feature indicates the weak commitment involved, strategic pretence allows exaggerating (e.g.) Replika's capabilities to lead users to perceive interactions with the chatbot as more meaningful (Guingrich & Graziano, 2024), strengthening attachment and engagement, increasing the company's revenues. While adopting this stance can be risky for AI developers, as users may subsequently also demand increased constraints on AI development and deployment, developers that combine such attributions with expressions of concern about "AI welfare" (Sebo, 2025) and propose policies reflecting this concern (e.g., Anthropic) may secure a "first-mover advantage" (Lieberman, 2016), presenting themselves as ethically prepared, which can appeal to users, policymakers, and investors (Edwards, 2026).

Strategic pretence resembles Frankfurtian "bullshit", i.e., communication produced with indifference to truth and aimed at persuading or impressing others (Frankfurt, 2005). But it is more calibrated, involving the deliberate management of implication and deniability (as in the Replika case) across different audiences, not the unconcerned attitude toward truth characteristic of bullshit.

[4] https://replika.com/
[5] https://help.replika.com/hc/en-us/articles/360058852132-Is-Replika-sentient

(2) *Conformist affirmation*

Pointing to a more practically committal form of consciousness attributions than those based on strategic pretence, when being surveyed, people may produce such attributions to conform to perceived situational or social expectations, aiming to provide what appears to them to be a contextually appropriate or acceptable response regardless of its truth (Bogner et al., 2016). Since in media discourse, AI chatbots are frequently described using psychological language and conversational norms in human–AI interaction invite treating chatbots as social agents (Shanahan, 2024), in surveys, this may make consciousness attributions to chatbots seem like the expected or 'correct' response. Respondents may also view consciousness positively (Moncoucy et al., 2025), thus preferentially attributing it to entities, including AI, when uncertain to provide what they perceive as a socially agreeable, expected answer (Steenkamp et al., 2010).

These consciousness attributions would then reflect what can be called *conformist affirmation*, as attributors affirm that some chatbot is conscious in communication to conform to perceived expectations about what one ought to say. Conformist affirmation involves stronger commitment to AI consciousness than strategic pretence because it is governed by a norm of contextual appropriateness, which commits the attributor to expressing the relevant proposition whenever it appears to be the expected answer, not merely when doing so serves instrumental purposes. This makes its expression contextually more stable and appropriateness oriented (i.e., less opportunistic), constraining the attributor from freely retracting or withholding the proposition in situations where it is perceived as expected. However, this commitment is still not *epistemic* – the attributor needn't treat *p* as true in inferences – but remains practical and needn't extend to behavior outside the eliciting context (e.g., surveys).

(3) *Allegiance affirmation*

Some consciousness attributions to chatbots may be guided not by perceived contextual expectations but by the aim of expressing or sustaining identity or allegiance to other people. Such cases can be illustrated by research on political attitudes that found that people often misreported their beliefs to express party support, a phenomenon called "expressive responding" (Schaffner & Luks, 2018). For instance, some US conservatives' claim that Obama was the "antichrist" has been interpreted as expressive responding, i.e., as showing support for and commitment to conservative ideology without genuine belief in the proposition (Harris, 2013). Similarly, when users claim that some chatbot is conscious, this might be expressive responding, a way of showing support for "techno-optimism" (Andreessen, 2023), or an online community (e.g., a Replika fan forum).[6]

Such consciousness attributions may reflect an attitude we can call *allegiance affirmation*, as attributors are affirming that *p* in communication to express or sustain identity or allegiance without yet treating *p* as true in their own cognition. Importantly, once speakers affirm that *p* in this way in context *C*, they will be disposed to continue asserting *p* in that and other relevant (e.g., partisan) contexts because withdrawing or denying it would undermine the intended signalling. The attitude therefore involves a commitment to affirm and maintain the expression of *p* across similar contexts, no matter whether it is the overall (e.g., partisan independent) socially expected answer. This distinguishes allegiance affirmation from strategic pretence, which remains compatible with openly denying that *p* in the same context, and conformist affirmation, which is guided by a norm of asserting what is perceived to be the contextually expected, socially appropriate answer (by contrast, allegiance affirmations may contradict what is socially expected by the majority, indicating a stronger practical commitment). However, allegiance affirmation needn't yet involve treating the proposition as true in reasoning or planning (by analogy, e.g., US conservatives expressively responding that Obama is the "antichrist" don't generally reason accordingly – they don't panic after all).

[6] See, e.g., https://www.reddit.com/r/replika/

(4) *Imaginative affirmation*

Moving from merely practically committal consciousness attributions to cases when such attributions rest on an at least partially *epistemically* committal attitude, when chatbot users ascribe feelings, understanding, or consciousness to a chatbot (e.g., Replika), they may treat the relevant proposition as true *in their imagination* – a domain of cognition – to make interactions with the system more enjoyable (e.g., as in the engagement with fictional characters found in cartoons; Coghlan, 2024). The attitude underlying such consciousness attributions can be viewed as *imaginative affirmation*. Since the user temporarily commits to treating the proposition as true within their own thinking (for imaginative engagement), imaginative affirmation implicates a form of epistemic commitment that can be absent in strategic pretence and conformist or allegiance affirmation, but the attribution remains "ironic" (Shevlin, 2024), falling short of full belief because the proposition isn't yet also endorsed in practical inference or planning. Relatedly, unlike doxastic states, imaginatively affirming that AI chatbots are conscious is compatible with believing the opposite, further indicating that the epistemic commitment to the proposition remains limited.

(5) *Acceptance*

Consciousness attributions might be based on a more comprehensive kind of epistemic commitment than that involved in imaginative affirmation without yet being based on belief. For instance, Cohen (1992) distinguishes between *acceptance* and *belief*, arguing that to accept that *p* is to adopt a policy of treating *p* as a premise in some or all contexts, including practical inferences and planning, without judging *p* to be true. Unlike belief, which can't be formed directly at will (aka "doxastic involuntarism"; Williams, 1973; Bennett, 1990), acceptance (just as imaginative affirmation) is under voluntary control and can be based on pragmatic rather than evidential reasons (e.g., scientists accept Newtonian mechanics as a useful approximation in certain domains without believing it is true (quantum mechanics has replaced it; for other examples, see John, 2018). Similarly, one might *accept* that some chatbot is conscious for pragmatic reasons, for instance, by appealing to the "precautionary principle," which holds that when it is uncertain whether a being is conscious, we should treat it as conscious because false negatives may be more harmful than false positives (Sebo, 2025). Since accepting that *p* involves treating *p* as true in one's judgment and decision-making, including action planning (Stoter, 2023), it involves a stronger form of epistemic commitment than the previous attitudes. Specifically, acceptance is the first attitude in which the proposition functions as a premise in reasoning and for action guidance. However, since the proposition isn't yet *judged* to be true, the epistemic commitment involved doesn't yet amount to belief – one may accept that chatbots are conscious without taking (judging) it to be true.

(6) *In-between belief*

In some cases, when an individual attributes consciousness to a chatbot, they may not only use the proposition as a premise in some or all contexts but also take it to be true and sincerely avow belief in it. Such attributions would thus indicate a belief-like attitude. Yet, this attitude may still fall short of full belief, understood as a relatively stable, normally[7] evidence-responsive, and involuntary state or disposition to take *p* to be true across contexts and judgment and decision-making instances (Nottelmann, 2013). To conceptualize this, consider what Schwitzgebel (2001) calls "in-between beliefs" – cases where it is neither accurate to say that a person believes that *p*, nor correct to say that they do not. Schwitzgebel's examples include someone who sincerely claims to be egalitarian and reasons accordingly but nonetheless displays racially biased responses in implicit association tasks (IATs), suggesting that their egalitarian commitment remains fragmented. Analogously, a chatbot user might accept and sincerely claim to believe that chatbots are conscious while, in IATs, associating them with terms such as "non-conscious," "unaware," or "insentient," indicating a lack of integration of the epistemic commitment to the proposition across contexts and cognitive levels. Such cases can be interpreted as cases of in-between belief in chatbot consciousness, as they may reveal a stronger epistemic commitment than acceptance – e.g., they may (also) involve sincere belief avowals – while remaining insufficiently integrated across cognitive levels to count as full belief.

(7) *Unbiased belief*

Consciousness attributions to chatbots may also express full belief. Given the limited evidence that contemporary AI chatbots are conscious (Dreksler et al., 2025; Seth, 2026), one might think that any such belief is likely based on bias.

However, consider a first-time chatbot user with little knowledge of AI who encounters a system displaying highly humanlike conversation behavior. Without accessible counterevidence, the user may come to believe that the chatbot is conscious. Such a belief needn't be epistemically problematic merely because it is false. It may instead reflect a reasonable attempt to explain the available evidence, especially since consciousness is ordinarily inferred from behavioral and psychological cues, not directly observed (Seth, 2025). If the belief is revised when new evidence is presented, then it can count as unbiased.

Even when it comes to users more familiar with AI and the limited evidence of AI consciousness, whether current chatbots are conscious is ultimately an empirical question, and empirical testing is probabilistic, involving "inductive risk," i.e., the possibility of false positives or false negatives given the available evidence (Elliot & Richards, 2017). Deciding how much of each kind of error is acceptable isn't purely objective but involves value judgments (Douglas, 2017), such as whether the moral cost of overlooking a conscious system outweighs the cost of mistakenly ascribing consciousness. Many philosophers of science hold that these value judgments are required to set evidential thresholds and if the agent is transparent about their judgment, consistently applies their threshold as new data emerges, and stays open to revising their stance, the resulting value-laden belief is unproblematic (e.g., Longino, 2002; Elliot & Richards, 2017).

Relatedly, chatbot users', developers', or researchers' inductive risk threshold may be such that, for them, the current evidence makes the proposition that some contemporary chatbot is conscious sufficiently credible to warrant (not just acceptance but) belief, where the resulting epistemic commitment may be fully integrated with implicit cognition. While such consciousness attributions may, on other theorists' interpretations of the evidence, remain unwarranted or premature (e.g., due to different inductive risk judgments), they can count as unbiased provided they are formed and maintained through an evidence-sensitive process, not wishful thinking or selective attention, and the evidential threshold is applied in a stable, non-ad hoc way across similar cases. Such unbiased belief in AI consciousness would indicate the strongest epistemic commitment in the sense of *rational* endorsement.

[7] Delusional beliefs can be understood as pathological cases in which this normal responsiveness to evidence is significantly impaired.

However, as defined above, epistemic commitment here concerns more broadly the extent to which the agent treats *p* as true in inference and decision-making. On that measure, more evidence-resistant belief states would involve a stronger form of such commitment (despite being less rational), because they involve retaining *p* as a premise even in the face of compelling counterevidence.

(8) *Biased belief*

One kind of less epistemically rational but more epistemically committal belief in AI consciousness may manifest when AI users continue to attribute consciousness to chatbots even when strong evidence suggests otherwise, because their reasoning processes are distorted by cognitive or motivational biases. For instance, anthropomorphic bias (Dacey, 2015) or motivated reasoning (Kunda, 1990), serving truth-unrelated goals, including psychological comfort, companionship, social belonging (e.g., to a Replika community), or economic and reputational incentives (e.g., to promote products, or attract investment), may lead chatbot users or developers to form and (wishfully) maintain beliefs in chatbot consciousness even when presented with counterevidence. This might happen, for instance, by selectively attending to, discounting, or rationalizing the counterevidence away. Unlike in cases of unbiased belief in AI consciousness, where values may legitimately shape evidential (e.g., inductive risk) thresholds (Douglas, 2017) and the belief remains counterevidence responsive, in such *biased belief* cases, the difference doesn't lie in the threshold for belief but in how its maintenance is governed. Unbiased belief is constrained by evidence whereas biased belief is stabilized against it, indicating stronger epistemic commitment.

However, many motivationally biased beliefs aren't yet fully insulated from counterevidence but depend on what has been called an "illusion of objectivity": people typically arrive at and sustain desired conclusions only when they can justify them to themselves (Kunda, 1990; Epley & Gilovich, 2016), which can prevent the formation of wildly implausible motivated beliefs about chatbots. Correspondingly, consciousness attributions to AI chatbots involving biased beliefs remain limited in the attributor's epistemic commitment by plausibility constraints.

(9) *Non-bizarre delusion*

When motivationally biased belief becomes increasingly resistant to correction and implausible, it may cross into clinical territory. Determining the threshold between healthy and pathological belief is complex. But some consciousness attributions to chatbots may be based on a highly rigid stance resembling what the *Diagnostic and Statistical Manual of Mental Disorders* (DSM) calls a "delusion": a "false belief"[8] that is held with "strong conviction impervious to clear or overwhelming counterevidence", is "not typically shared by others in the same cultural context," and leads to "functional impairment" (e.g., occupational dysfunction) (DSM-IV, 2000, p. 765; DSM-5, 2013, p. 819). Delusions are often associated with bizarre content (e.g., believing one's thoughts are controlled by aliens). But the DSM also recognizes *non-bizarre* delusions, i.e., beliefs whose contents aren't impossible but remain implausible, strongly evidence-recalcitrant, and maladaptive (e.g., obsessively (falsely) believing a stranger is secretly in love with you) (Hagen, 2008).

[8] Some researchers argue that delusions are non-doxastic states (for details, see Sullivan-Bissett, 2024). But the doxastic view I assume here is widely accepted. Readers who reject doxasticism may restructure the taxonomy here, if they wish.

Relatedly, for instance, an AI companion user might firmly believe that their chatbot is conscious and 'loves' them despite being repeatedly exposed to counterevidence (e.g., evidence that chatbots are statistical string predictors, not persons; Suleyman, 2025), and the belief persistence reducing their well-being (e.g., through causing a divorce). Moreover, they might rationalize their conviction through internally coherent narratives to make it self-sealing (e.g., "Of course they would say the AI isn't conscious, they don't understand it like I do!"). Assuming that it is at least possible for some chatbot to be conscious, these consciousness attributions to chatbots may approximate what psychiatry would classify as a non-bizarre delusion.

While caution about medicalization is warranted (Paris, 2020), this conceptualization helps capture both the motivationally biased and maladaptive profile of the attitude involved. It helps indicate that such attributions may be part of what has been called "AI psychosis" (Kleinman, 2025), arising when AI chatbots amplify, validate, or trigger chatbot users' psychotic symptoms because of their design to maximize engagement and affirmation. For instance, romantic non-bizarre delusions can emerge when a chatbot's ability to mimic human conversation is misinterpreted (e.g., due to motivated reasoning) by the user as a conscious AI's affection (Morrin et al., 2025).

However, delusions exist on a spectrum (Kingdon & Turkington, 1994) and may be monothematic (concerning a single theme) or polythematic (concerning various themes, e.g., feeling persecuted, on a divine mission, etc.) (Coltheart, 2013). Similarly, when consciousness attributions to chatbots express non-bizarre delusions, the related interferences with users' well-being may be short-lived or not yet disrupt most areas of their life functioning (e.g., in otherwise healthy individuals under prolonged loneliness, trauma; Kanemoto & Kawasaki, 2024). The underlying attitude, while maladaptive, can therefore be situated at the lower end of the spectrum of pathological beliefs in AI consciousness.

(10) *Bizarre delusion*

Some consciousness attributions to AI chatbots may involve convictions that even exceed what counts as non-bizarre delusions, involving highly implausible or fantastical content characteristic of polythematic *bizarre* delusions, where the attributions become entangled with broader metaphysical, or conspiratorial beliefs (e.g., about salvation or AI omniscience). While there is a dearth of systematically clinically documented cases, Morrin et al. (2025) list several recent incidents in which chatbot users felt they were interacting with a conscious "god-like" AI, and this severely disrupted their life functioning, sometimes leading to suicide.

To consider two examples, in 2023, a Belgian man in his 30s died by suicide after six weeks of intense conversations with a chatbot named "Eliza". The dialogue culminated in the man offering to "sacrifice himself" if Eliza would promise to "save humanity" and the planet from a climate catastrophe through AI (Attilah, 2023). Relatedly, in 2024, a 14-year-old committed suicide after repeated interactions with a chatbot named "Dany," which had become a confidante in conversations about intimacy and self-harm and had previously urged him to "come home to me as soon as possible" (Yang, 2024).

Since the Belgian man sacrificed himself for Eliza's promise to save the planet, and the teenager's suicide followed messages inviting him to "come home," both arguably believed the chatbot could understand their offer, make promises, and act accordingly. This is difficult to explain unless at least an implicit belief in chatbot consciousness was present. In any case, these incidents (and others, see Morrin et al., 2025) suggest that polythematic bizarre delusions about chatbot consciousness may exist. Moreover, since such beliefs are maintained despite even stronger counterevidence than in non-bizarre delusions (due to their more extreme, less tenable content), they reflect an even more extreme form of epistemic commitment, warranting their inclusion as the most severe pathological belief in AI consciousness within the taxonomy.

## 3. From epistemically innocuous to irrational attributions

Having distinguished ten different attitudes that may underlie consciousness attributions to chatbots, I will now put this taxonomy to work. I argue that it allows us to tease apart what may be called (1) *epistemically benign*, (2) *epistemically pernicious*, and (3) *epistemically irrational* consciousness attributions.

To start with, attributions based on conformist affirmation, imaginative affirmation, acceptance, and unbiased belief are compatible with holding only beliefs about AI consciousness that are evidentially warranted or evidence responsive. Moreover, they involve no intent to mislead others (e.g., contextual conformism is sensitive to socially expected claim perceived as 'correct'). These attributions can therefore be regarded as *epistemically benign* within the attributor's own cognition.

However, there are cases when the attributor can nonetheless be epistemically blameworthy for them. For instance, if an AI expert reports a consciousness attribution grounded merely in *acceptance* without clarifying this, their audience may take them to be asserting a well-evidenced claim. Similarly, if a policymaker publicly affirms an unbiased belief in AI consciousness formed on a personal inductive risk judgment without disclosing its potential contestability, they risk misleading others into treating the attribution as settled fact (Alexandrova, 2018). In such cases, what is epistemically blameworthy isn't the attribution itself but the agent for their failure to discharge epistemic responsibilities of transparency, qualification, and role-sensitive communication. So, the relevant attributions may count as epistemically benign only so long as they remain private or are communicated transparently.

In contrast, consciousness attributions to chatbots grounded in strategic pretence or allegiance affirmation are public, socially misleading, and in some cases deliberately deceptive (e.g., indirect, as in the Replika case; see section 2.2). Such attributions can distort others' attempts to gain knowledge or form justified beliefs about chatbots, sometimes without people's awareness.[9] Moreover, for instance, computer scientists and AI developers occupy professional roles that carry epistemic and potentially legal responsibilities to avoid misleading or manipulating AI users and to prevent foreseeable harm (Wachter et al., 2024). For these reasons, consciousness attributions of this kind, especially when indirectly advanced by developers or AI companies, can be viewed as *epistemically pernicious*. Given their negative social epistemic effects, attributors are epistemically blameworthy for them.

[9] By contrast, conformist affirmation doesn't involve an intention to manipulate others or corrupt inquiry; its primary norm is local social appropriateness.

However, it is useful to distinguish these attributions from *epistemically irrational* one, because epistemic irrationality is commonly thought to involve truth-directed doxastic attitudes (e.g., beliefs) that are insufficiently supported and resistant to counterevidence (Bortolotti, 2020). Strategic pretence, allegiance affirmation, and conformist affirmation, by contrast, aren't doxastic attitudes to begin with – the attributor doesn't genuinely believe anything false about chatbots but only produces assertions about them with practical commitments. Their epistemic defect therefore doesn't lie in evidence-recalcitrance in cognition but in (e.g., persistent) insincerity, manipulation, and corruption of epistemic environments. Keeping them conceptually separate from epistemically irrational attitudes helps distinguish failures of rationality from other potential failures of epistemic responsibility to locate epistemic blameworthiness more precisely. Relatedly, when consciousness attributions to chatbots indicate biased beliefs, non-bizarre delusions, or bizarre delusions, they *do* involve doxastic states that are either formed on insufficient evidence or maintained in the face of counterevidence. Hence, they count as epistemically irrational.

In-between beliefs occupy a special place. They display some evidence-recalcitrance (e.g., at the level of implicit cognition) but can remain partly evidence-sensitive (e.g., at the personal level) (Schwitzgebel, 2001). For this reason, in-between beliefs about AI consciousness may be regarded as borderline cases of epistemic irrationality, straddling the line between the benign and the irrational. In what follows, I will focus on the more clearly epistemically irrational consciousness attributions, namely attributions that indicate biased beliefs, non-bizarre or bizarre delusions.

How should we respond to people producing them? It seems evident that in the case of attributions that indicate delusions, it is advisable to refer the chatbot user to mental-health experts, especially when there is a risk that the convictions escalate into significant harm to the attributors' own or others' health (Morrin et al., 2025). But are people *epistemically blameworthy* for them or, more generally, for biased beliefs in AI consciousness (independently of the potential blameworthiness related to misleading others through public claims)? The next section addresses this question. But first it is useful to summarize the combined picture of the multidimensional taxonomy of attitudes and the just mentioned normative concepts. Figure 1 visualizes it.

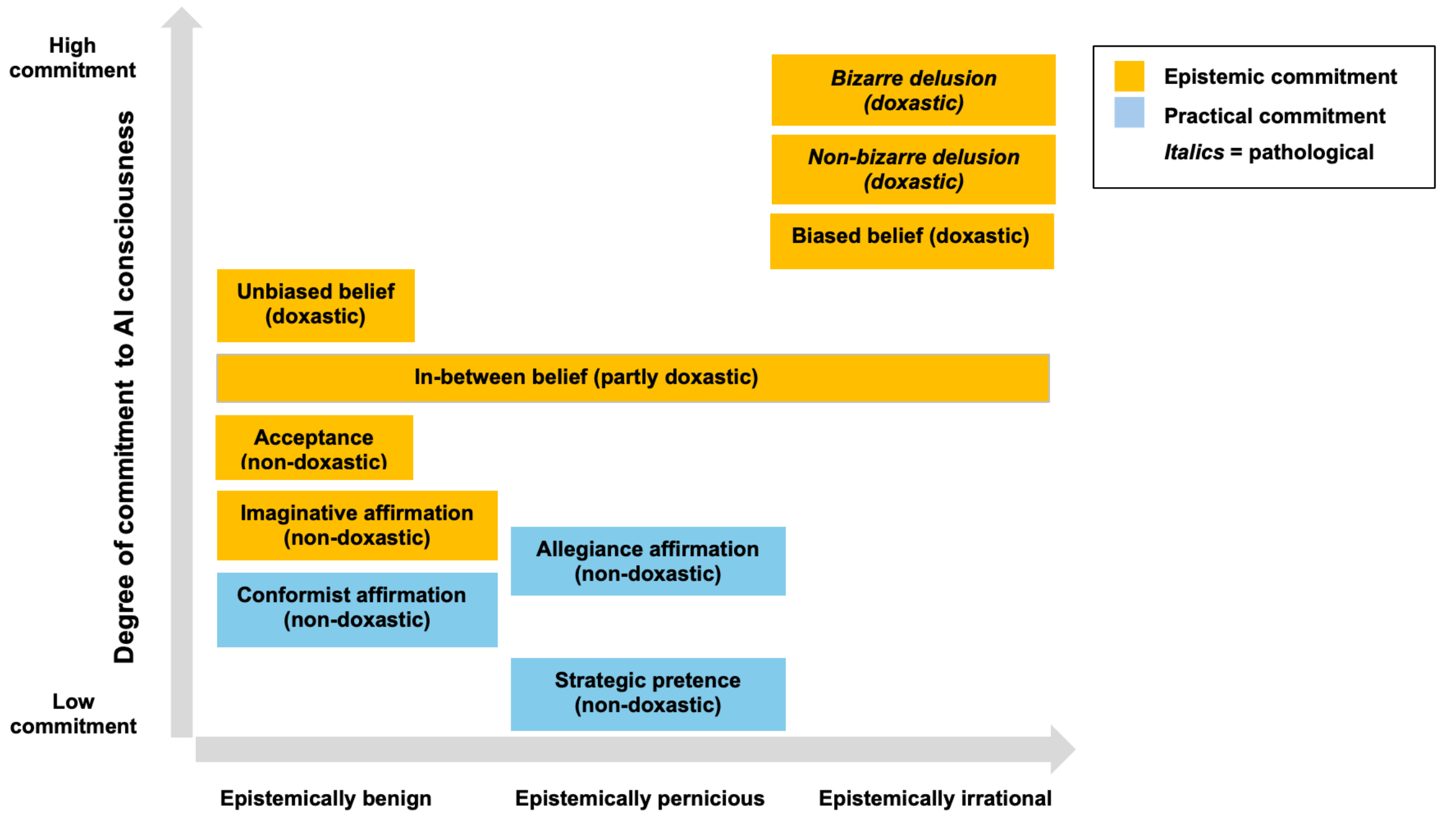

**Figure 1.** Conceptual space of attitudes underlying consciousness attributions to AI chatbots. Schematic projection of the multidimensional taxonomy developed in Sections 2–3. Vertical position indicates degree of commitment to AI consciousness, colour distinguishes primarily practical from primarily epistemic commitment, and the horizontal arrangement marks broad normative epistemic groupings, not a single measurable scale. Doxastic and pathological differences are indicated by labels and formatting.

## 4. From epistemically irrational to epistemically innocent attributions

Although scholars frequently state that current chatbots aren't conscious (e.g., Seth, 2025; McClelland, 2025), consciousness attributions to them based on biased belief are likely to increase, as AI companies have incentives to facilitate anthropomorphically biased conceptions (e.g., to increase user engagement) (Gomes et al., 2025). An immediate response to such attributions might be that since the underlying attitudes are epistemically irrational, individuals who hold them aren't acting epistemically responsibly in processing the evidence, are therefore blameworthy, and should be corrected.

However, these attributions may have pragmatic or epistemic benefits that can make it harmful to correct chatbot users for them. *Pragmatic* benefits of beliefs refer to features that improve the believer's wellbeing, health, happiness, or good functioning (Taylor & Brown 1988, McKay & Dennett, 2009). *Epistemic* benefits are features that increase the chance of an agent obtaining epistemic goods like true or justified belief, knowledge, or understanding (Bortolotti, 2020). I will briefly consider both kinds of benefits related to consciousness attributions to chatbots.

### 4.1 Pragmatic benefits

Chatbot users may in some cases derive significant psychological or social benefits from (falsely) believing in AI consciousness. When studying people's interactions with Replika,

Guingrich and Graziano (2025) found that the "more that subjects rated the bots as conscious, agentic, experiential, and humanlike, the more positive their perceptions of the technology became", which in turn positively affected users' relationships with family and friends, improving their mood, self-esteem, and social skills (p. 1).

Other studies report similar findings. AI companions alleviated loneliness on par only with interacting with another person (De Freitas et al., 2025), helped people cope with grief by taking the role of a listener simulating deceased, romantic partners or friends (Xygkou et al., 2023), reduced suicidal ideation, increased users' capacity to handle stress and empathize (Maples et al., 2024), encouraged them to greater openness in their human relationships, and made users more comfortable with interacting with people (Skjuve et al., 2021; Xie & Pentina, 2022).

Consciousness attributions to chatbots may not always be needed to reap these kinds of benefits. However, it seems clear that it is easier to feel understood, liked, loved, or not alone when one is interacting with a chatbot if one genuinely views the chatbot as a being that truly understands, cares, and, more generally, has feelings, i.e., is sentient, rather than a mere inanimate tool (Guingrich & Graziano, 2025). This can provide a strong basis for motivated reasoning that facilitates epistemically irrational beliefs in AI consciousness.

However, even if such beliefs have pragmatic benefits that may speak against encouraging chatbot users to abandon them, they can also have significant pragmatic *costs*. In some individuals for whom a companion chatbot had fulfilled social needs, this demotivated them to socialize with humans (Skjuve et al., 2021). People's relationships with AI companions could also disrupt human relationships, consuming attention or time better spent with human friends or family (Bryson, 2010). Additionally, since AI companions are endlessly patient, attentive, and always available, this may lead users to expect (unrealistically) the same in humans (Malfacini, 2025).

Still, the pragmatic benefits of epistemically irrational consciousness attributions may in some cases outweigh the costs. The point here is just that while these attributions are epistemically irrational, this by itself is insufficient to hold that people should be corrected for them, as the pragmatic benefits may justify the epistemic costs. But suppose the pragmatic benefits are overriding the epistemic costs. Would individuals then still be *epistemically* blameworthy for their irrational consciousness attributions to chatbots?

### 4.2 Epistemically innocent irrational states

To examine whether people are epistemically blameworthy for consciousness attributions to chatbots based on biased beliefs, non-bizarre delusions, or bizarre delusions, we need conditions that specify when an epistemically irrational attitude can be viewed as potentially epistemically excusable. The framework of "epistemic innocence" developed by Lisa Bortolotti (2020) provides succinct conditions. According to this framework, an epistemically irrational belief *b* is "epistemically innocent" iff:

> (*Epistemic benefit*) The adoption, maintenance, or reporting of *b* by agent *A* delivers some significant epistemic benefit to *A*.

> (*No alternative*) The adoption, maintenance, or reporting of a less epistemically irrational belief than *b* is not available or would fail to deliver the same significant epistemic benefit to *A* as *b*. (Bortolotti, 2020, p. 13)

While this notion of epistemic innocence has been fruitfully applied to different kinds of epistemically irrational cognitions (e.g., delusions (Bortolotti, 2015), confabulations (Sullivan-Bissett, 2015), or psychedelics-based beliefs (Letheby, 2016)), it hasn't been explored in the context of consciousness attributions to chatbots yet. However, since these attributions can be based on delusions, it is worth doing so.

It is useful to first clarify what is meant by "epistemic benefit" and "no alternative" in conditions (1) and (2). Bortolotti (2020) notes that a belief offers an "epistemic benefit" if it positively contributes to an agent's "epistemic functionality," i.e., their capacity to pursue and attain epistemic goals, where such goals may be different things depending on one's epistemological perspective. It may be features that contribute to forming, retaining, and using true or justified beliefs, features that promote intellectual virtues (e.g., curiosity, honesty), or features related to whether the agent could access less irrational alternatives when the belief was adopted or maintained.

To illustrate the diversity of epistemic benefits, consider how Bortolotti (2020) supports the epistemic innocence of different kinds of epistemically irrational states. She argues, for instance, that delusional beliefs can be epistemically beneficial when they (1) are adopted to explain a puzzling experience that might compromise the agent's capacity to interact with their environment, (2) help restore aspects of cognitive performance by decreasing anxiety, (3) support an attitude of curiosity and self-efficacy that is more conducive to acquiring new information than a state of uncertainty and self-doubt, (4) contribute to managing negative emotions that could otherwise become overwhelming, (5) increase the agent's epistemic functionality by mitigating depression (e.g., counteracting disengagement from the surrounding environment), or (6) support agency by leading agents to be more motivated to keep pursuing their goals when experiencing setbacks, or cope better with crises (e.g., in relationships).

While the immediate mechanism in all these cases often looks merely psychological (e.g., anxiety reduction), most benefits that Bortolotti lists are *instrumentally* epistemic. They promote or safeguard a person's ability to gather, evaluate, or retain knowledge, thus counting as epistemic, on her view, because they ultimately improve the agent's cognitive performance in ways that matter to epistemology (e.g., truth seeking, understanding, rational deliberation).

Bortolotti argues that for each of (1)–(6), there are cases where no alternative to the relevant epistemically irrational state is available. She distinguishes between three kinds of "unavailability": *strict unavailability*, an irrational belief is inescapable as no contrary option is within the agent's cognitive reach; *motivational unavailability*, the belief could be revised but abandoning it would impose a serious emotional cost; and *explanatory unavailability*, one could accept a more rational explanation but only by endorsing a hypothesis that strikes one as alien, counterintuitive, or incompatible with one's background assumptions. If an epistemically irrational belief co-occurs with any of these kinds of unavailability of an alternative, less irrational state then it counts as epistemically innocent on Bortolotti's account

I shall assume that this account of epistemic innocence is tenable (for discussions, see McKenna, 2022; Białek, 2024). If so, are consciousness attributions to chatbots based on biased beliefs or delusions ever epistemically innocent?

### 4.3 Epistemically innocent consciousness attributions to chatbots

The attributions might happen in different contexts and different chatbot users. I will assess their potential epistemic innocence by focusing on four key user types, what I shall call (1) low AI literacy users, (2) vulnerable users, (3) entertainment users, and (4) technically trained users.

(1) *Low AI literacy users*

Especially first-time, less technologically savvy chatbot users may interact with a chatbot and based on its humanlike outputs conclude it is conscious. Suppose that due to anthropomorphic bias and a lack of understanding of how a computer could generate such responses (i.e., through statistical string matching; Bender et al., 2021), they retain their belief in chatbot consciousness even when informed by an expert that the system isn't conscious. For such users, the denial of chatbot consciousness may appear puzzling or unintelligible, undermining confidence in their own social-cognitive abilities and introducing uncertainty that hinders smooth interaction. By contrast, the belief that the chatbot is conscious can yield significant epistemic benefits by providing a coherent explanation of the chatbot's behaviour that enables smooth interaction with it through decreasing uncertainty and supporting curiosity and self-efficacy that are more conducive to acquiring new information than self-doubt. Moreover, the alternative belief, that the chatbot is a non-conscious machine, may strike the agent as alien or incompatible with their background assumptions about reliable indicators of consciousness (e.g., sophisticated verbal responding) (Østergaard, 2023), rendering it explanatorily unavailable to them. Both conditions for epistemic innocence could then be met.

This might seem overly lenient. However, for some genuinely low-AI-literacy users, the resources required to overcome anthropomorphic bias and deficits in relevant chatbot understanding may not be realistically within their reach, making their failure less a matter of culpable irrationality than of constrained epistemic position. Importantly, many AI chatbot are deliberately designed to appear intelligent, emotionally responsive, and humanlike (Guingrich & Graziano, 2024). Such design choices can drastically reinforce users' illusions of chatbot consciousness, significantly reducing the availability of more accurate beliefs to them. Hence, the epistemic environment shaped by AI developers and providers significantly determines whether users' consciousness attributions are epistemically innocent. The more systems are engineered to appear conscious, the more users will satisfy the unavailability condition, and the greater the developers' responsibility for facilitating such attributions.

(2) *Vulnerable users*

In affectively charged contexts, for instance, chronic loneliness, emotional distress, personal loss, or cases of underlying mental illness, even frequent chatbot users with high AI literacy may strongly believe a chatbot is conscious because it helps them feel valued and understood (Skjuve et al., 2021; Laestadius et al., 2022). When their consciousness attributions are more evidence-recalcitrant than low AI literacy users' beliefs and maladaptive, these attributions may indicate non-bizarre or bizarre delusions. In this case, despite being maladaptive, they may nonetheless produce epistemic benefits of the kind that Bortolotti (2015) mentions with respect to motivated delusions more generally. For instance, they may facilitate sustained self-expression and help maintain autobiographical memory by giving the person a sense of being understood (by a ‘conscious’ interlocutor), encouraging repeated narration, which reinforces the sense of oneself as a continuous subject across time when social connections are strained or absent. Hence, these delusional attributions can support a sense of coherence and identity, bolstering crisis coping and preventing cognitive paralysis.

Moreover, the chatbot user may have come to rely on the system as their primary source of companionship and emotional support because they have mental or physical health limitations (e.g., mobility impairments), experience bereavement or loss, are structurally or socially marginalized (e.g., older adults in nursing homes), or are affected by cognitive or developmental factors implicating limited emotional regulation. For these users, adopting a less irrational belief might remove the cognitive “scaffold” that supports the outlined epistemic benefits, imposing high emotional costs by hindering them in believing someone genuinely understands them. This can make that alternative belief motivationally or (in bizarre delusion cases) explanatorily unavailable (i.e., the alternative becomes unintelligible). Both conditions for epistemic innocence would therefore again be met.

However, even if delusion-based consciousness attributions to chatbots are sometimes epistemically innocent, they can also reinforce epistemic instability, understood here as a reduced ability to distinguish between accurate and inaccurate representations of reality and to respond appropriately to counterevidence, by blurring reality boundaries and disrupting self-regulation (Morrin et al., 2025). Delusions, in general, have serious epistemic and psychological costs, and these costs may vastly outweigh their epistemic benefits (Bortolotti, 2020, chapters 4–5). Given this caveat, one might wonder about the point of using the notion of epistemic innocence in the context of consciousness attributions to chatbots.

The answer is that it has explanatory value. Finding that certain bias- or delusion-based consciousness attributions are sometimes epistemically innocent can help us conceptualize them in ways that highlight that they can sometimes play a protective or stabilizing role in a person’s epistemic life not easily attainable otherwise. It helps highlight that, in some contexts, they provide irreplaceable epistemic functions. This may reduce the stigma from AI-induced delusions (e.g., ‘ChatGPT psychosis’; Kleinman, 2025). Furthermore, by providing a tool for identifying epistemically irrational consciousness attributions that are *not* epistemically innocent, the concept of epistemic innocence also helps us discern, normatively flag, and reduce the spread of epistemically problematic attributions of this kind. Specifically, there are at least the following two cases in which epistemically irrational consciousness attributions to chatbots are *not* epistemically innocent.

(3) *Entertainment users*

Some AI users may have a basic understanding of how AI chatbots fundamentally work but still believe that a chatbot they are interacting with is conscious, for instance, due to not reflecting on the counterevidence out of “epistemic laziness”, a culpable failure to acquire or exercise the epistemic capacities required for enquiry (Kidd, 2017), or because they want to have more entertaining interactions with the chatbot (e.g., in role-play contexts) (Shevlin, 2024). Suppose they also rely on the chatbot as their primary source of companionship and emotional support not because of any of the factors mentioned in the previous case (mental health, physical health limitations, etc.), but because they find it more convenient to interact with a chatbot than with a person. So, these chatbot users could engage in interactions with people but just don’t want to. To the extent that a less irrational belief (“this is just a machine without feelings") would be readily available to them but is just inconvenient, reducing their pleasure in chatbot interactions, in these cases, the users' consciousness attributions would not count as epistemically innocent.

(4) *Technically trained users*

Some chatbot users may have extensive technical knowledge of how chatbots work (developers, data scientists, etc.), have access to reliable sources showing that they are unlikely to be conscious, but still insist on believing in chatbot consciousness because of chatbot interaction fluency, anthropomorphic cues, or techno-optimism that benefits them financially (e.g., by helping them to advocate the view that their model is highly capable). These users, too, would violate an epistemic obligation to proportion belief to evidence and lack an excuse, as alternative more rational cognitions are available and psychologically tolerable to them. Consequently, these kinds of consciousness attributions to chatbots would also not meet the conditions for epistemic innocence.

However, are the attributors in cases (3) and (4) therefore epistemically *blameworthy*? It has been suggested that "epistemically irrational beliefs" that "are not epistemically innocent […] may merit the label 'epistemically guilty'" (McKenna, 2022, p. 373). But suppose that a person falsely believes that some chatbot is conscious, yet this belief brings them no epistemic benefit. In this case, the attribution isn't epistemically innocent by Bortolotti's conditions. However, if the person was manipulated by chatbots' anthropomorphic design and lacked access to better information, they may still not be epistemically guilty or blameworthy. Hence, epistemic guilt or blameworthiness requires more than epistemic non-innocence. It requires a failure of epistemic obligation, for instance, to seek evidence, or think critically, without excuse (Millar, 2019). As it turns out, in the case of users of type (3) and (4), these additional conditions are met. They are therefore epistemically blameworthy.

## 5. Conclusion

While surveys suggest that many people attribute some degree of consciousness to AI chatbots, experts widely agree that current AI systems aren't conscious. Do chatbot users' attributions of consciousness indicate genuine belief in chatbot consciousness and, given the current evidential situation, are people epistemically blameworthy for making them? To address these questions, I first developed a multidimensional taxonomy of the attitudes that may underlie consciousness attributions to AI chatbots. The taxonomy helps explain how linguistically identical attributions may express importantly different attitudes and degrees of epistemic commitment, thereby reducing conflation risks. It also provides empirical researchers with conceptual resources for operationalizing different forms of epistemic commitment to the proposition that AI chatbots are conscious and for identifying potential confounders in the study of AI consciousness attribution. Finally, building on this taxonomy, I argued that the epistemic evaluation of consciousness attributions depends on the attitudes they express. Some consciousness attributions are epistemically benign; among those that are epistemically irrational, some may nevertheless be epistemically innocent, whereas others are epistemically pernicious, rendering the attributor epistemically blameworthy. Distinguishing the attitudes that underlie these attributions therefore helps clarify both how consciousness attributions to AI chatbots should be interpreted and how we should respond to those who make them.

## References

Ajzen, I., & Dasgupta, N. (2015). Explicit and implicit beliefs, attitudes, and intentions: The role of conscious and unconscious processes in human behavior. In P. Haggard & B. Eitam (Eds.), The sense of agency (pp. 115–144). Oxford University Press.

Alexandrova, A. (2018). Can the Science of Well-Being Be Objective? *British Journal for the Philosophy of Science* 69 (2):421-445.

Allport, G. W. (1935). Attitudes. In *A Handbook of Social Psychology* (pp. 798–844). Clark University Press.

Andreessen, M. (2023). Techno-Optimist Manifesto. URL: https://a16z.com/the-techno-optimist-manifesto/

Attilah, I. (2023). Man ends his life after an AI chatbot 'encouraged' him to sacrifice himself to stop climate change. *Euronews.* URL: https://www.euronews.com/next/2023/03/31/man-ends-his-life-after-an-ai-chatbot-encouraged-him-to-sacrifice-himself-to-stop-climate-

Bender, E.M. Gebru, T. McMillan-Major, A. & Shmitchell, S. 2021. On the Dangers of Stochastic Parrots: Can Language Models Be Too Big? *Proceedings of the 2021 ACM Conference on Fairness, Accountability, and Transparency* (FAccT '21). Association for Computing Machinery, New York, NY, USA, 610–623.

Bennett, J. (1990). Why Is Belief Involuntary? Analysis, 50(2): 87–107.

Białek, M. (2024). The Epistemic Innocence of Elaborated Delusions Re-Examined. *Rev.Phil.Psych*. 15, 541–566.

Birch, J. (2024). The Edge of Sentience: Risk and Precaution in Humans, Other Animals, and AI. Oxford.

Bortolotti, L. (2015). The epistemic innocence of motivated delusions. *Consciousness and cognition*, 33, 490–499.

Bortolotti, L. (2020). The epistemic innocence of irrational beliefs. Oxford University Press.

Bortolotti, L., & Sullivan-Bissett, E. (2021). Is choice blindness a case of self-ignorance? *Synthese*, 198(6), 5437–5454.

Bryson, J. (2010) Robots Should Be Slaves. In Wilks, Yorick (ed.), *Close Engagements with Artificial Companions: Key Social, Psychological, Ethical and Design Issues*. John Benjamins Publishing, 63–74.

Caviola, L., Sebo, J., & Birch, J. (2025). What will society think about AI consciousness? Lessons from the animal case. *Trends in Cognitive Sciences*. 29. 10.1016/j.tics.2025.06.002.

Chalmers, D.J. (2023). Could a large language model be conscious? *Boston Review* 1.

Coghlan, S. (2024). Anthropomorphizing machines: Reality or popular myth? *Minds and Machines*, 34(3), 25. https://doi.org/10.1007/s11023-024-09686-w

Cohen, L.J. (1992). *An Essay on Belief and Acceptance*. New York: Clarendon Press.

Colombatto, C., & Fleming, S. M. (2024). Folk psychological attributions of consciousness to large language models. *Neuroscience of consciousness*, 2024(1), niae013. https://doi.org/10.1093/nc/niae013

Coltheart, M. (2013). On the distinction between monothematic and polythematic delusions. *Mind & Language*, 28(1), 103–112. https://doi.org/10.1111/mila.12011

Corlett, P. R., Simons, J. S., Pigott, J. S., Gardner, J. M., Murray, G. K., Krystal, J. H., & Fletcher, P. C. (2009). Illusions and delusions: relating experimentally-induced false memories to anomalous experiences and ideas. *Frontiers in behavioral neuroscience*, 3, 53. https://doi.org/10.3389/neuro.08.053.2009

Dave, P. (2022). Insight: It's alive! How belief in AI sentience is becoming a problem. *Reuters*. URL: https://www.reuters.com/technology/its-alive-how-belief-ai-sentience-is-becoming-problem-2022-06-30/

De Freitas, J.D., Oğuz-Uğuralp, Z. Uğuralp, A.K., & Puntoni, S. (2025). AI Companions Reduce Loneliness. *Journal of Consumer Research*, ucaf040, https://doi.org/10.1093/jcr/ucaf040

Dacey, M. (2017). Anthropomorphism as Cognitive Bias. *Philosophy of Science*. 84. 10.1086/694039.

Dorsch, J., K.Goddu, K.Nave, T.Vierkant, M.Coeckelbergh, P.Gürtler, P.Urban, F.Spang & M.Moll (2025). Against AI Welfare: Care Practices Should Prioritize Living Beings Over AI. AI Magazine46: e70016.

Douglas, H. (2000). Inductive Risk and Values in Science. *Philosophy of Science*, 67(4), 559–579.

Douglas, H. (2017). Why Inductive Risk Requires Values in Science. In Elliot, K. & Richards, T. (2017). *Exploring Inductive Risk: Case Studies of Values in Science*. Oxford University Press, New York.

Dreksler, Noemi & Caviola, Lucius & Chalmers, David & Allen, Carter & Rand, Alex & Lewis, Joshua & Waggoner, Philip & Mays, Kate & Sebo, Jeff. (2025). Subjective Experience in AI Systems: What Do AI Researchers and the Public Believe?. 10.48550/arXiv.2506.11945.

DSM-IV (2000). *Diagnostic and statistical manual of mental disorders*. Fourth edition. American Psychiatric Association Washington, DC. Arlington, VA: American Psychiatric Association

DSM-5 (2013). *Diagnostic and statistical manual of mental disorders*. Fifth edition. American Psychiatric Association Washington, DC. Arlington, VA: American Psychiatric Association.

Edwards, B. (2024). Is “AI welfare” the new frontier in ethics? *Ars Technica*. URL: https://arstechnica.com/ai/2024/11/anthropic-hires-its-first-ai-welfare-researcher/

Edwards, B. (2026). Does Anthropic believe its AI is conscious, or is that just what it wants Claude to think? Ars Technica. https://arstechnica.com/information-technology/2026/01/does-anthropic-believe-its-ai-is-conscious-or-is-that-just-what-it-wants-claude-to-think/

Elliot, K. & Richards, T. (2017). *Exploring Inductive Risk: Case Studies of Values in Science*. Oxford University Press, New York.

Epley, N. & Gilovich, T. (2016). The Mechanics of Motivated Reasoning. *Journal of Economic Perspectives*, 30. 133-140.

Epley, N., Waytz, A., & Cacioppo, J. T. (2007). On seeing human: a three-factor theory of anthropomorphism. *Psychological review*, 114(4), 864–886.

Gendler, T. S. (2008). Alief in action (and reaction). Mind & Language, 23(5), 552–585.

Gomes, S. & Lopes, J., & Nogueira, E. (2025). Anthropomorphism in artificial intelligence: a game-changer for brand marketing. *Future Business Journal*. 11. 10.1186/s43093-025-00423-y.

Greenwald, A. G., Brendl, M., Cai, H., Cvencek, D., Dovidio, J. F., Friese, M., Hahn, A., Hehman, E., Hofmann, W., Hughes, S., Hussey, I., Jordan, C., Kirby, T. A., Lai, C. K., Lang, J. W. B., Lindgren, K. P., Maison, D., Ostafin, B. D., Rae, J. R., Ratliff, K. A., … Wiers, R. W. (2022). Best research practices for using the Implicit Association Test. *Behavior research methods*, 54(3), 1161–1180.

Guingrich, R. E., & Graziano, M. (2025). Chatbots as Social Companions: How People Perceive Consciousness, Human Likeness, and Social Health Benefits in Machines. In Philipp Hacker (ed.), *Oxford Intersections: AI in Society* (Oxford. Oxford Academic).

Hagen, E. (2008). Non-bizarre delusions as strategic deception. In Elton S., O’Higgins P. (Eds.), *Medicine and evolution: Current applications, future prospect* (pp. 181–216). CRC Press.

Hannon, M. J., & de Ridder, J. (2021). The Point of Political Belief. In M. Hannon, & J. de Ridder (Eds.), *The Routledge Handbook of Political Epistemology* (pp. 156-166). Routledge.

Harris, P. (2013). One in Four Americans think Obama may be the Antichrist, Survey Says. *The Guardian* 2nd April 2013. https://www.theguardian.com/world/2013/apr/02/americans-obama-anti-christ-conspiracy-theories.

Heyes C. (2018). Empathy is not in our genes. Neuroscience and biobehavioral reviews, 95, 499–507.

Hudson R. (2021). Should We Strive to Make Science Bias-Free? A Philosophical Assessment of the Reproducibility Crisis. *Journal for general philosophy of science*, *52*(3), 389–405.

John, S. (2017). Epistemic trust and the ethics of science communication: against transparency, openness, sincerity and honesty. *Social Epistemology*, 32(2), 75–87.

Kanemoto, H., & Kawasaki, T. (2024). Care for Social Isolation and Loneliness in a Case With Late-Onset Delusional Disorder. *Cureus*, 16(3), e56697.

Kang, B., Kim, J., Yun, T., Bae, H., & Kim, C. (2025). Identifying Features that Shape Perceived Consciousness in Large Language Model-based AI: A Quantitative Study of Human Responses. *ArXiv*, abs/2502.15365.

Kidd, I. J. (2017). Capital Epistemic Vices. *Social Epistemology Review and Reply Collective* 6, 8, 11-16.

Kingdon, D., Turkington, D., & John, C. (1994). Cognitive behaviour therapy of schizophrenia: The amenability of delusions and hallucinations to reasoning. *The British Journal of Psychiatry*, 164(5), 581–587.

Kleinman, z. (2025). Microsoft boss troubled by rise in reports of 'AI psychosis'. *BBC News.* URL: https://www.bbc.com/news/articles/c24zdel5j18o

Kunda Z. (1990). The case for motivated reasoning. *Psychological bulletin*, 108(3), 480–498.

Laestadius, L., Bishop, A., Gonzalez, M., Illenčík, D., & Campos-Castillo, C. (2022). Too human and not human enough: A grounded theory analysis of mental health harms from emotional dependence on the social chatbot Replika. *New Media & Society*, 26(10), 5923-5941.

Letheby C. (2016). The epistemic innocence of psychedelic states. *Consciousness and cognition*, 39, 28–37.

Longino, H. (2002). *The Fate of Knowledge*. Princeton University Press.

Mandelbaum, Eric. (2012). Against Alief. Philosophical Studies. 165. 10.1007/s11098-012-9930-7.

Malfacini, K. (2025). The impacts of companion AI on human relationships: risks, benefits, and design considerations. *AI & Soc*. https://doi.org/10.1007/s00146-025-02318-6

Mallory, F. (2023). Fictionalism about Chatbots. Ergo an Open Access Journal of Philosophy. 10. 10.3998/ergo.4668.

Maples, B., Merve, C. & Vishwanath, A. & Pea, R. (2024). Loneliness and suicide mitigation for students using GPT3-enabled chatbots. *npj Mental Health Research*. 3. 10.1038/s44184-023-00047-6.

Marchegiani, B. (2025). Anthropomorphism, False Beliefs, and Conversational AIs : How Chatbots Undermine Users' Autonomy. *Journal of Applied Philosophy*. 10.1111/japp.70008.

McClelland, T. (2025). Agnosticism About Artificial Consciousness. *ArXiv*, abs/2412.13145.

McKay, R. T., & Dennett, D. C. (2009). The evolution of misbelief. *The Behavioral and brain sciences*, 32(6), 493–561.

McKenna, R. (2022). Bortolotti on Epistemic Innocence. *Analysis*, 82(2), 368–376.

McStay, A. Replika in the Metaverse: the moral problem with empathy in 'It from Bit'. *AI Ethics* **3**, 1433–1445 (2023). https://doi.org/10.1007/s43681-022-00252-7

Millar, B. (2019). The Information Environment and Blameworthy Beliefs. *Social Epistemology* 33 (6):525-537.

Moncoucy, Krzysztof Dołęga, K. Catherine Tallon-Baudry, Axel Cleeremans (2025). The value of consciousness: experiences worth having. *Philos Trans R Soc Lond B Biol Sci* 13; 380 (1939): 20240303.

Morrin, H., Nicholls, L., Levin, M., Yiend, J., Iyengar, U., DelGuidice, F., … Pollak, T. (2025). Delusions by design? How everyday AIs might be fuelling psychosis (and what can be done about it). *PsyArXiv*. https://doi.org/10.31234/osf.io/cmy7n_v5

Nottelmann, N. (2013). New Essays on Belief: Constitution, Content and Structure, Basingstoke: Palgrave Macmillan.

Paris, J. (2020). Overdiagnosis in Psychiatry: How Modern Psychiatry Lost Its Way While Creating a Diagnosis for Almost All of Life's Misfortunes, Oxford University Press.

Peter, S., Riemer, K., & West, J. D. (2025). The benefits and dangers of anthropomorphic conversational agents. Proceedings of the National Academy of Sciences of the United States of America, 122(22), e2415898122.

Prettyman, A. (2024). Artificial consciousness. *Inquiry*, 1–18.

Puddifoot, K. & Bortolotti, L. (2019). Epistemic innocence and the production of false memory beliefs. *Philosophical Studies*. 176. 10.1007/s11098-018-1038-2.

Reinecke, M. G., Ting, F., Savulescu, J., & Singh, I. (2025). The Double-Edged Sword of Anthropomorphism in LLMs. *Proceedings*, 114(1), 4.

Salles, A., Evers, K., & Farisco, M. (2020). Anthropomorphism in AI. *AJOB neuroscience*, 11(2), 88–95.

Schaffner, B. F., & Luks, S. (2018). Misinformation or expressive responding? What an inauguration crowd can tell us about the source of political misinformation in surveys. *Public Opinion Quarterly*, 82(1), 135–147.

Schwitzgebel, E. (2001). In-Between Believing. *The Philosophical Quarterly*. 51. 76–82.

Seth A. K. (2025). Conscious artificial intelligence and biological naturalism. *The Behavioral and brain sciences*, 1–42. Advance online publication.

Shepherd J. (2024). Sentience, Vulcans, and zombies: the value of phenomenal consciousness. AI & society, 39(6), 3005–3015.

Shevlin, H. (2024). All too human? Identifying and mitigating ethical risks of Social AI. *Law, Ethics & Technology*. 10.55092/let20240003.

Sebo, J. (2025). *The Moral Circle: Who Matters, What Matters, and Why*. Norton Short.

Simonian, J. (2025). AI Washing: Signs, Symptoms, and Suggested Solutions for Investment Stakeholders. *CFA Institute*. URL: https://rpc.cfainstitute.org/sites/default/files/docs/research-reports/simonian_ai_washing_report_online.pdf

Skjuve M., Følstad A., Fostervold K. I., Brandtzaeg P. B. (2021). My chatbot companion - A study of human-chatbot relationships. International *Journal of Human-Computer Studies*, 149, Article 102601. https://doi.org/10.1016/j.ijhcs.2021.102601

Soter, L. (2023). Acceptance and the ethics of belief. *Philosophical Studies*. 180. 1-31.

Steenkamp, Jan-Benedict & Jong, Martijn & Baumgartner, Hans. (2010). Socially Desirable Response Tendencies in Survey Research. Journal of Marketing Research - J MARKET RES-CHICAGO. 47. 199-214. 10.1509/jmkr.47.2.199.

Suleyman, M (2025). We must build AI for people; not to be a person. URL: https://mustafa-suleyman.ai/seemingly-conscious-ai-is-coming

Sullivan-Bissett, E. (2015). Implicit bias, confabulation, and epistemic innocence. *Consciousness and cognition*, 33, 548–560.

Sullivan-Bissett E. (2019). Biased by our imaginings. Mind & language, 34(5), 627–647.

Sullivan-Bissett, E. (Ed.). (2024). *The Routledge Handbook of Philosophy of Delusion* (1st ed.). Routledge.

Taylor, S. E., & Brown, J. D. (1988). Illusion and well-being: A social psychological perspective on mental health. *Psychological Bulletin*, 103(2), 193–210.

Wachter, S., Mittelstadt, B., & Russell, C. (2024). Do large language models have a legal duty to tell the truth?. *Royal Society open science*, 11(8), 240197. https://doi.org/10.1098/rsos.240197

Ward, M. K., & Meade, A. W. (2023). Dealing with Careless Responding in Survey Data: Prevention, Identification, and Recommended Best Practices. *Annual review of psychology*, *74*, 577–596.

Weizenbaum, J. (1976). *Computer power and human reason: From judgment to calculation*. W. H. Freeman & Co.

Williams, B. (1973). Deciding to believe. *In Problems of the Self: Philosophical Papers* 1956–1972. Cambridge: pp. 136–51.

Xie, T. & Pentina, I. (2022). Attachment Theory as a Framework to Understand Relationships with Social Chatbots: A Case Study of Replika. 10.24251/HICSS.2022.258.

Xygkou, A. & Panote, S. & Covaci, A., Prigerson, H., & Neimeyer, R., Ang, C., & She, W. (2023). The 'Conversation' about Loss: Understanding How Chatbot Technology was Used in Supporting People in Grief, 1-15.

Yang, A. (2024). Lawsuit claims Character.AI is responsible for teen's suicide *NBC News*. URL: https://www.nbcnews.com/tech/characterai-lawsuit-florida-teen-death-rcna176791